\documentclass[twocolumn,showpacs,preprintnumbers,amsmath,amssymb]{revtex4}
\usepackage{graphicx}
\usepackage{epsfig}
\usepackage{dcolumn}
\usepackage{bm}

\newcommand{\be}{\begin{enumerate}}
\newcommand{\ee}{\end{enumerate}}
\newcommand{\bi}{\begin{itemize}}
\newcommand{\ei}{\end{itemize}}

\def\Journal#1&#2&#3(#4){#1{\bf #2}, #3 (#4)}

\def\bec{\begin{center}}
\def\eec{\end{center}}

\begin{document}
\title{\Large \boldmath Partial wave analyses of $J/\psi\to\gamma\pi^+\pi^-$ and
$\gamma\pi^0\pi^0$}

\author{
M.~Ablikim$^{1}$,              J.~Z.~Bai$^{1}$,               Y.~Ban$^{12}$,
J.~G.~Bian$^{1}$,              X.~Cai$^{1}$,                  H.~F.~Chen$^{17}$,
H.~S.~Chen$^{1}$,              H.~X.~Chen$^{1}$,              J.~C.~Chen$^{1}$,
Jin~Chen$^{1}$
% ...
\\
\vspace{0.2cm}
(BES Collaboration)\\
\vspace{0.2cm}
{\it
$^{1}$ Institute of High Energy Physics, Beijing 100049, People's Republic of China\\
% ...
$^{12}$ Peking University, Beijing 100871, People's Republic of China\\
% ...
$^{17}$ University of Science and Technology of China, Hefei 230026, People's Republic of China\\
% ...
}
} 

\noindent\vskip 0.2cm 
\begin{abstract}
  Results are presented on $J/\psi $ radiative decays to $\pi^+\pi^-$
  and $\pi^0\pi^0$ based on a sample of 58M $J/\psi$ events taken with
  the BES\,II detector.  Partial wave analyses are carried out using
  the relativistic covariant tensor amplitude method in the 1.0 to 2.3
  GeV/$c^2$ $\pi \pi$ mass range.  There are conspicuous peaks due
  to the $f_2(1270)$ and two $0^{++}$ states in the 1.45 and 1.75
  GeV/$c^2$ mass regions. The first $0^{++}$ state has a mass of
  $1466\pm 6\pm 20$ MeV/$c^2$, a width of $108{^{+14}_{-11}}\pm 25$
  MeV/$c^2$, and a branching fraction ${\cal B}(J/\psi\to \gamma f_0(1500)
  \to\gamma \pi^+\pi^-) = (0.67\pm0.02\pm0.30) \times 10^{-4}$.
  Spin 0 is strongly preferred over spin 2. The second $0^{++}$ state peaks
  at $1765^{+4}_{-3}\pm 13 $ MeV/$c^2$ with a width of $ 145\pm8\pm69
  $ MeV/$c^2$. If this $0^{++}$ is interpreted as coming from $f_0(1710)$, 
  the ratio of its branching fractions to $\pi\pi$ and $K\bar K$ is
  $0.41^{+0.11}_{-0.17}$.
\end{abstract}

\pacs{12.39.Mk, 13.25.Gv, 14.40.Cs}

\maketitle 

\section{\boldmath Introduction}   \label{introd} 

QCD predicts the existence of glueballs, the bound states of gluons,
and the observation of glueballs would provide a direct test of
QCD. In the quenched approximation, lattice QCD calculations predict
the lightest glueball to be a $0^{++}$ with the mass being in the
region from 1.4 to 1.8 GeV/$c^2$~\cite{QCDL}. Although the
identification of a glueball is very complicated, there are several
glueball candidates, including the $f_0(1500)$ and $f_0(1710)$. The
properties of the $f_0(1500)$ and $f_0(1710)$ are reviewed in detail
in the latest issue of the Particle Data Group (PDG)~\cite{PDG}.

$J/\psi$ radiative decays have been suggested as promising modes
for glueball searches. The $J/\psi\to\gamma\pi^+\pi^-$ process was
analyzed previously in the Mark\,III~\cite{mark3}, DM2~\cite{dm2} 
and BES\,I~\cite{bes1charge} experiments, in which there was evidence for $f_2(1270)$
and an additional $f_2(1720)$. However, the high mass shoulder of the
$f_2(1270)$, at about 1.45 GeV/$c^2$, was unsettled. A revised
amplitude analysis of Mark\,III data assigned the shoulder to be a
scalar at $\sim$ 1.43 GeV/$c^2$, and, in addition, found the peak at
$\sim$ 1.7  GeV/$c^2$ to be scalar rather than tensor~\cite{dunwoodie}. The
$J/\psi\to\gamma\pi^0\pi^0$ process was also studied by the
Crystal Ball~\cite{cball} and BES\,I experiments~\cite{bes1}, but no
partial wave analysis has yet been performed on this channel. In this
paper, the results of partial wave analyses on
$J/\psi\to\gamma\pi^+\pi^-$ and $\gamma\pi^0\pi^0$ are presented based
on a sample of 58M $J/\psi$ events collected by the upgraded Beijing
Spectrometer (BES\,II) located at the Beijing Electron Positron Collider
(BEPC).

\section{\boldmath BES detector}
BES\,II is a large solid-angle magnetic spectrometer that is described
in detail in Ref.  ~\cite{detector}. Charged particle momenta are
determined with a resolution of $\sigma_p/p$ = 1.78 \% $\sqrt{1+p^2}$
($p$ in GeV/$c$) in a 40-layer cylindrical main drift chamber
(MDC). Particle identification is accomplished by specific ionization
($dE/dx$) measurements in the drift chamber and time-of-flight (TOF)
measurements in a barrel-like array of 48 scintillation counters. The
$dE/dx$ resolution is $\sigma_{dE/dx}$ = 8.0 \%; the TOF resolution is
$\sigma_{TOF} = 180$ ps for Bhabha events. Outside of the
time-of-flight counters is a 12-radiation-length barrel shower counter
(BSC) comprised of gas tubes operating in limited stream mode.  The
BSC measures the energies of photons with a resolution of
$\sigma_E/E\simeq$ 21 \% /$\sqrt{E}$ ($E$ in GeV). Outside the
solenoidal coil, which provides a 0.4 T magnetic field over the
tracking volume, is an iron flux return that is instrumented with
three double layers of\ counters that are used to identify muons.

In this analysis, a GEANT3 based Monte Carlo simulation program
(SIMBES)~\cite{simbes} with detailed consideration of detector
performance (such as dead electronic channels) is used.  The
consistency between data and Monte Carlo has been checked in many high
purity physics channels, and the agreement is quite
reasonable~\cite{simbes}.


\section{\boldmath Event selection}

The first level of event selection of $J/\psi\to\gamma\pi^+\pi^-$
requires two charged tracks with total charge zero. Each charged
track, reconstructed using MDC information, is required to be well
fitted to a three-dimensional helix, be in the polar angle region
$|\cos\theta_{MDC}|<0.8$, and have the point of closest approach of
the track to the beam axis be within 2 cm of the beam axis and within
20 cm from the center of the interaction region along the beam line.

More than one photon per event is allowed because of the possibility
of fake photons coming from the interactions of charged tracks with
the shower counter or from electronic noise in the shower counter.  A
neutral cluster is considered to be a photon candidate when the energy
deposited in the BSC is greater than 50 MeV, the first hit is in the
beginning six radiation lengths, the angle between the nearest charged
track and the cluster is greater than 18$^{\circ}$, and the angle
between the cluster development direction in the BSC and the photon
emission direction is less than 30$^{\circ}$.

The total number of layers with hits associated with the two charged
particles in the muon counter is required to be less than four in
order to remove $\gamma\mu^+\mu^-$ events.  To remove the large
backgrounds from Bhabha events, we require that (i) the opening angle of
the two tracks satisfies $\theta_{op} < 175^{\circ}$  and (ii) the
energy deposit by each track in the BSC satisfies $E_{SC} < 1.0$
GeV. We require $\theta_{op} > 10^{\circ}$ to remove $\gamma$
conversions that occur at low $\pi^+\pi^-$ mass. In order to reduce
the background from final states with kaons and electrons, both tracks
are required to be identified as pions by TOF or $dE/dx$ when the
momenta are lower than 0.7 GeV/$c$.  In other cases, at least one
track is required to be identified as a pion by TOF.

Requirements on two variables, $U$ and $P^2_{t\gamma}$, are
imposed~\cite{THimel}. The variable $U=(E_{miss}-|\vec{P}_{miss}|)$ is
required to satisfy $|U|<0.15$ GeV. Here, $E_{miss}$ and
$\vec{P}_{miss}$ are, respectively, the missing energy and momentum of
charged particles. The variable
$P^2_{t\gamma}=4|\vec{P}_{miss}|^2\sin^2\theta_{\gamma}/2$ is required
to be $< 0.0045$ (GeV/$c$)$^2$, where $\theta_{\gamma}$ is the angle
between the missing momentum and the photon direction. The $U$ cut
removes most background from events having multikaon or other neutral
particles, such as $K^*(892)^{\pm}K^{\mp}$,~$\gamma K^+K^-$ events.
The cut on $P^2_{t\gamma}$ is used to reduce backgrounds with
$\pi^0$s.

In order to reduce the dominant $\rho\pi$ background, events with more
than one photon
satisfying $|M_{\gamma_1\gamma_2}-M_{\pi^0}|<0.065$ GeV/$c^2$ are
rejected.  Here $M_{\gamma_1\gamma_2}$ is the invariant mass of the
two isolated photons with the smallest angle between the plane
determined by these two photons and the direction of $\vec{P}_{miss}$
in all possible photon combinations.  $M_{\gamma_1\gamma_2}$ is
calculated using $P_{miss}$ and the angle between $\vec{P}_{miss}$ and
the $\gamma$ direction. The advantage of this method is that it uses
the momenta of the charged tracks measured by the MDC, which has good
momentum resolution, and is independent of photon energy measurement.

Finally, the two charged tracks and photon in the event are
kinematically fitted using four energy and momentum conservation
constraints (4-C) under the $J/\psi \to \gamma \pi^+ \pi^-$ hypothesis
to obtain better mass resolution and to suppress backgrounds further
by using the requirements $\chi ^2_{\gamma \pi^+\pi^-} < 15$ and $\chi
^2_{\gamma \pi^+\pi^-} < \chi^2_{\gamma K ^+K ^-}$. If there is more
than one photon, the fit is repeated using all permutations and the
combination with the best fit to $\gamma \pi^+\pi^-$ is retained.

For $J/\psi\to\gamma\pi^0\pi^0$, the $\pi^0$ mesons in the event are
identified through the decay $\pi^0\to\gamma\gamma$.  The isolated
photon is required to have the energy deposited in the BSC greater
than 80 MeV and come from the interaction point. The number of
isolated photons is required to be greater than four and less than
seven.  A 4-C kinematic fit to $J/\psi\to5\gamma$ is performed, the
combination of five photons with the smallest $\chi^2$ is selected,
and a kinematic fit Chi-square $\chi^2_{5\gamma} < 15$ is
required. For five photons, there are 15 combinations from which to
construct two $\pi^0$s. To select $\pi^0$s, we choose the combination
with the smallest $\Delta$, where $\Delta=\sqrt{(M_{\gamma_1
\gamma_2}-M_{\pi^0})^2+(M_{\gamma_3 \gamma_4}-M_{\pi^0})^2}$ and
require $|M_{\pi^0_{1,2}}-M_{\pi^0}|<40$ MeV/$c^2$.  To reduce
background with $\omega$s, events with the invariant mass of a $\pi^0$
and one photon in the $\omega$ mass interval
$|M_{\gamma\pi^0_{1(2)}}-M_{\omega}|<30$ MeV/$c^2$ are rejected. 
To further suppress backgrounds with more than one neutral particle recoling
to the $\pi^0\pi^0$ system, the recoiling mass squared of the $\pi^0\pi^0$
system is required to be less than 4.8 (GeV/$c^2$)$^2$.

\begin{figure}[htbp]
   \includegraphics[width=0.35\textwidth]
       {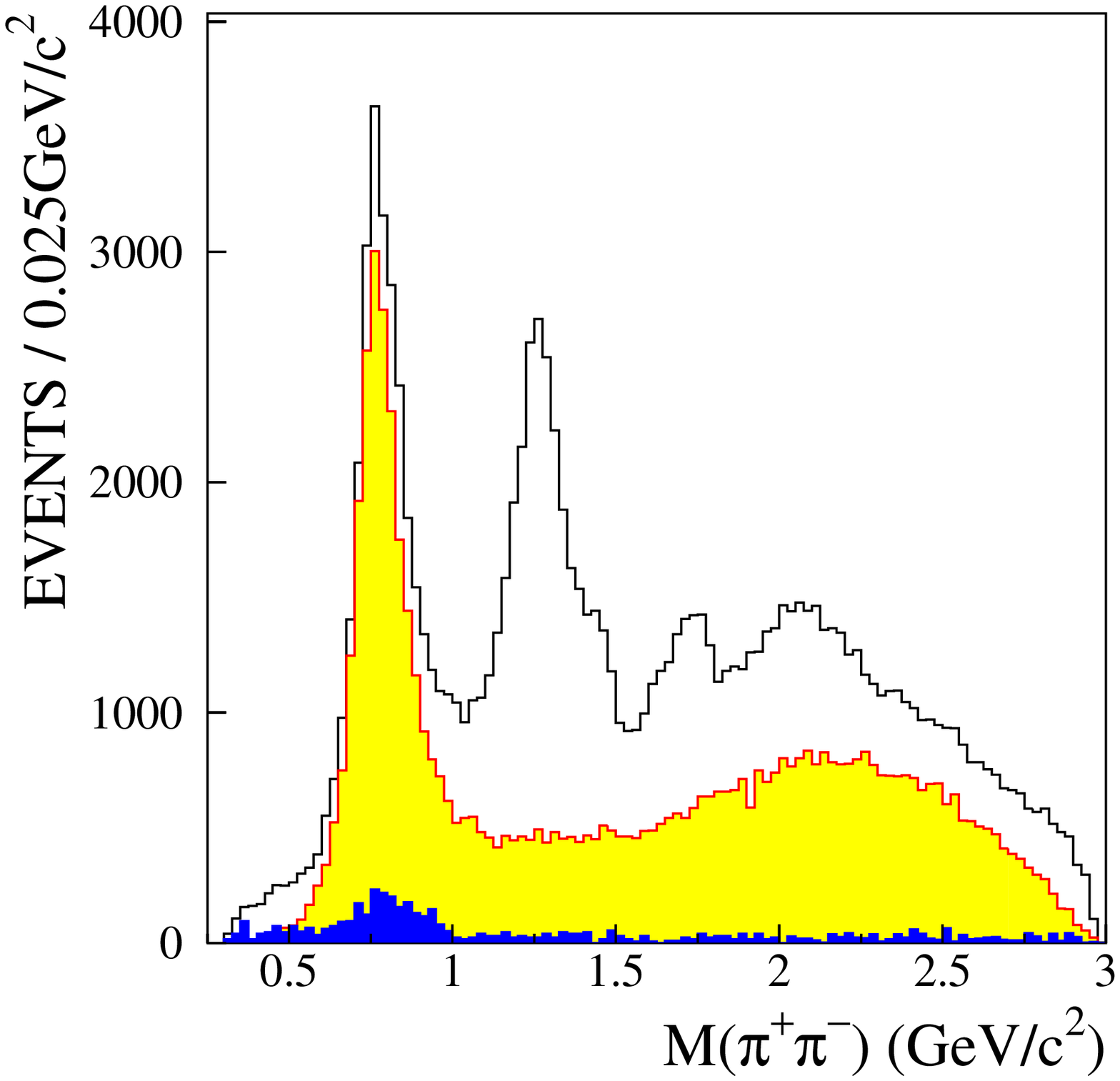}
   \vspace{1mm}
   \includegraphics[width=0.35\textwidth]
       {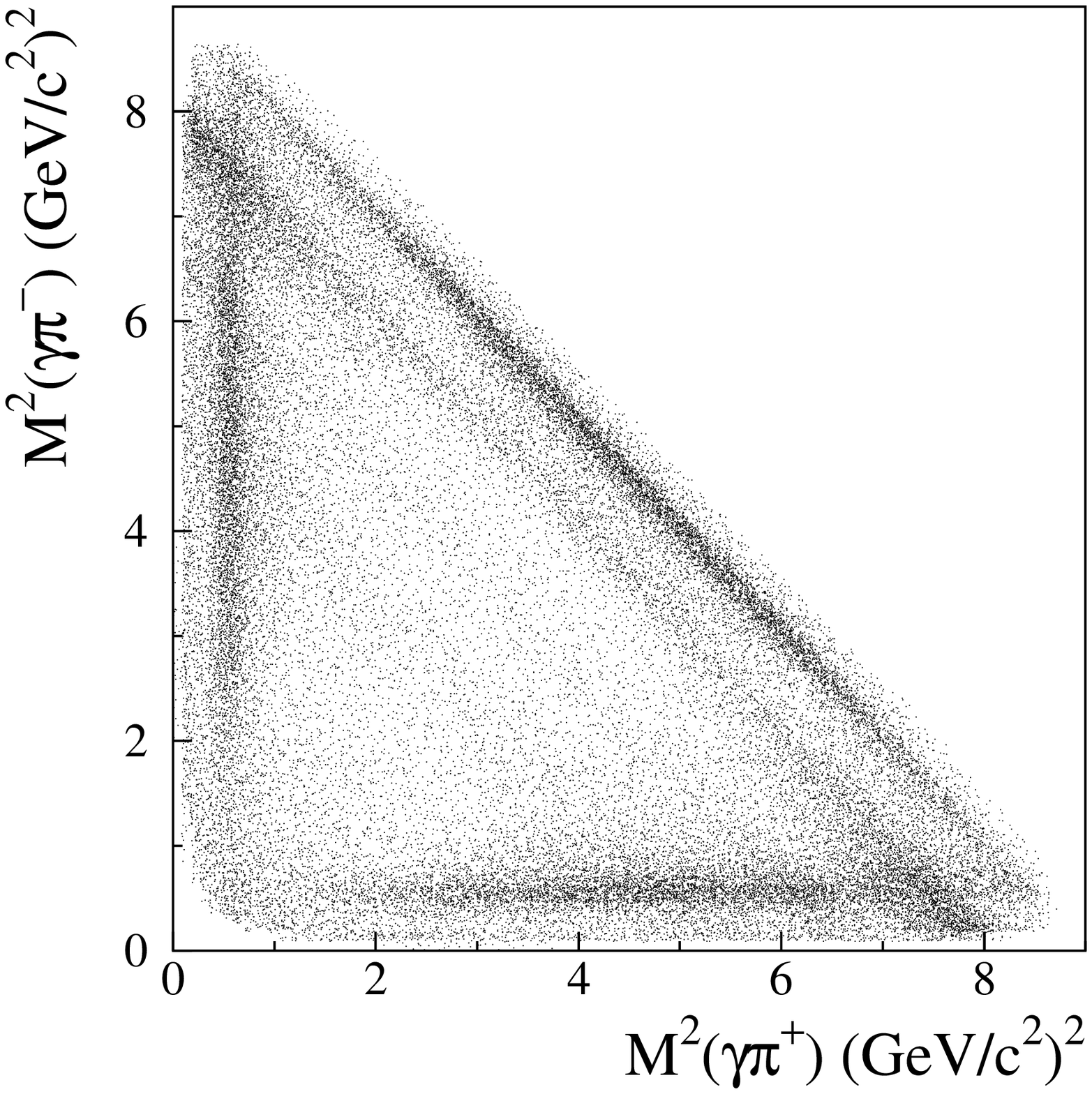}
   \caption{Invariant mass spectrum of $\pi^+\pi^-$ and the Dalitz
plot for $J/\psi\to\gamma\pi^+\pi^-$, where the lightly and dark shaded 
histograms in the upper panel correspond to $J/\psi\to\pi^+\pi^-\pi^0$
and other estimated backgrounds, respectively.}
   \label{mass1}
   \end{figure}

Fig. \ref{mass1} shows the $\pi^+\pi^-$ mass spectrum for the
selected events, together with the corresponding background 
distributions and the Dalitz plot.
There is a strong $\rho^0(770)$ peak mainly due to background from 
$J/\psi\to\rho^0\pi^0$. A strong $f_2(1270)$ signal, a shoulder on the 
high mass side of the $f_2(1270)$, an enhancement at $\sim$ 1.7 GeV/$c^2$, 
and a peak at $\sim$ 2.1 GeV/$c^2$ are clearly visible. 
The lightly shaded histogram in Fig. \ref{mass1} corresponds to the 
dominant background $J/\psi\to\pi^+\pi^-\pi^0$.
The data taken at the $e^+e^-$ center of mass energy of 3.07 GeV, with a
luminosity of $2272.8\pm 36.4$ nb$^{-1}$, are used to determine the continuum
background. The sum of continuum background and the other possible 
backgrounds, such as 
$J/\psi \to \gamma \eta^{\prime}$ ($\eta^{\prime}\to \gamma \rho^{0},
~\rho^{0} \to \pi^+ \pi^-$), $J/\psi\to K^*(892)^{\pm} K^{\mp}$, $\dots$, is estimated to be 
3.8 \% of the data in the whole mass range and is shown as the dark shaded 
histogram in Fig. \ref{mass1}.

Fig. 2 shows the $\pi^0\pi^0$ mass spectrum and the Dalitz plot. 
The shaded histogram corresponds to the sum of estimated backgrounds 
determined using PDG branching ratios~\cite{PDG}. The backgrounds 
are mainly from $J/\psi\to \omega \pi^0 \pi^0$ ($\omega \to \gamma \pi^0$, 
$\pi^0\to 2\gamma$), $J/\psi\to \gamma \eta$, ($\eta \to 3\pi^0,
~\pi^0 \to 2\gamma$) and $J/\psi \to \gamma\eta\pi^0\pi^0$ 
($\eta\to 2\gamma$, $\pi^0\to 2\gamma$). The peak below 0.5 GeV/$c^2$ 
is mainly from $J/\psi\to\gamma\eta$, ($\eta\to 3\pi^0,~\pi^0\to 2\gamma$).
The continuum background is also studied using the data taken at the
$e^+e^-$ center of mass energy of 3.07 GeV. No events survive selection
requirements. In general, the $\pi^+\pi^-$ and $\pi^0\pi^0$ 
mass spectra exhibit similar structures above 1.0 GeV/$c^2$.

   \begin{figure}[htbp]
   \includegraphics[width=0.35\textwidth]
       {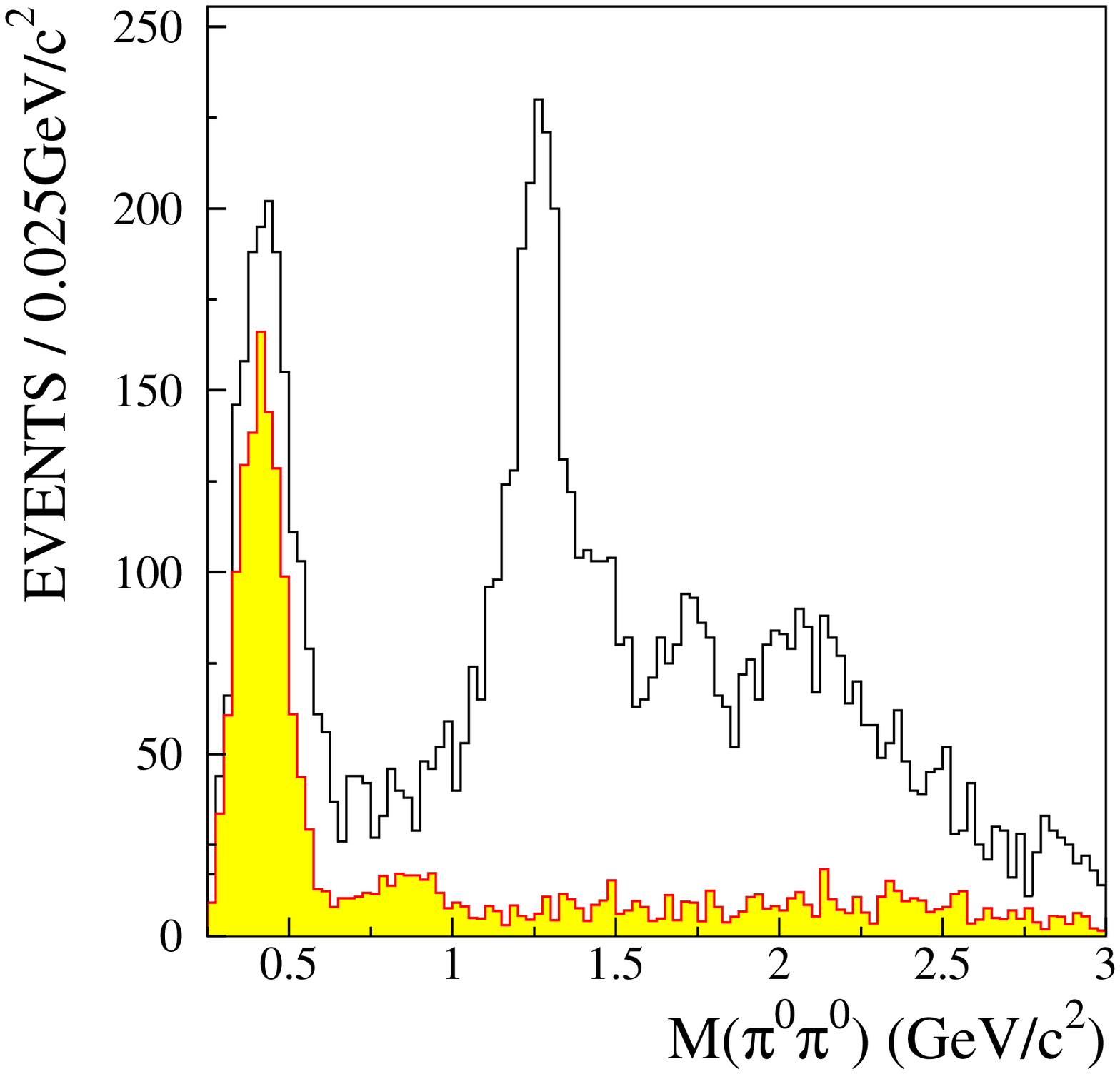}
   \vspace{1mm}
   \includegraphics[width=0.35\textwidth]
       {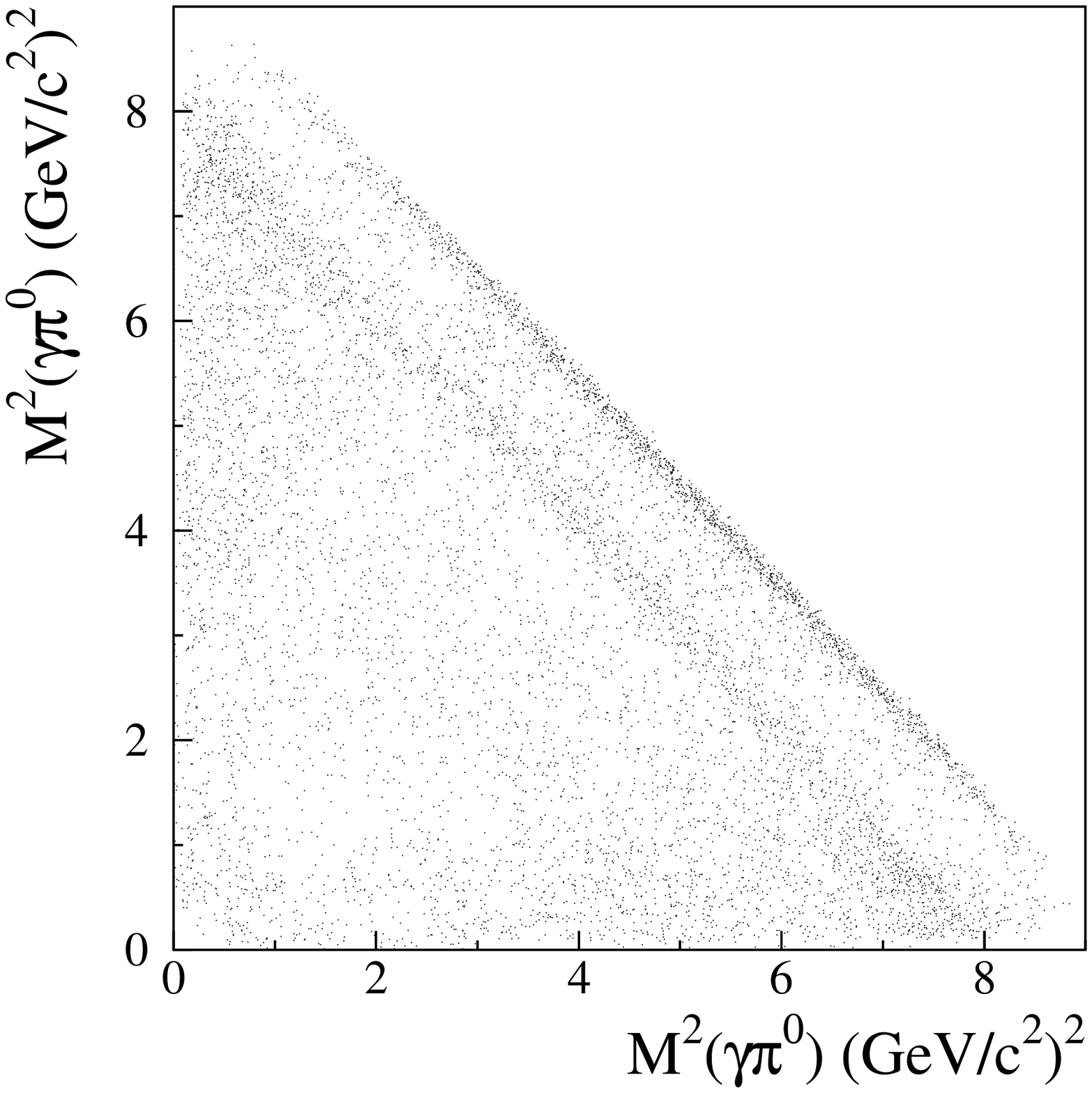}
   \caption{Invariant mass spectrum of $\pi^0\pi^0$ and the Dalitz
plot for $J/\psi\to\gamma\pi^0\pi^0$, where the shaded histogram
in the upper panel corresponds to the estimated backgrounds.}
   \label{mass2}
   \end{figure}

\vspace{1.5cm}
\section{\boldmath partial wave analysis}

We have carried out partial wave analyses for the $\pi\pi$ mass range from
1.0 to 2.3 GeV/$c^2$ using relativistic covariant
tensor amplitudes constructed from Lorentz-invariant combinations of
the polarization and momentum 4-vectors of the initial and final state
particles, with helicity $\pm 1$ for  $J/\psi$ initial states~\cite{pwa}.
Cross sections are summed over photon polarizations.  
The relative magnitudes and phases of the
amplitudes are determined by a maximum likelihood fit. 


For $J/\psi\to\gamma\pi^+\pi^-$, 
the following channels are fitted to the data:     
\begin {eqnarray*}
J/\psi &\to& \gamma f_2(1270) \\
       &\to& \gamma f_0(1500) \\
       &\to& \gamma f_0(1710) \\
       &\to& \gamma f_2(1810) \\
       &\to& \gamma f_0(2020) \\
\end {eqnarray*}
\begin {eqnarray*}
       &\to& \gamma f_2(2150) \\
       &\to& \gamma f_4(2050).
\end {eqnarray*}

Constant width Breit-Wigner functions are used for each
resonance. The form is described as follows:
$$
BW_{X}=\frac{m\Gamma}{s-m^2+im\Gamma},
$$
where $s$ is the square of $\pi^+\pi^-$ invariant mass, $m$ and $\Gamma$ are
the mass and width of intermediate resonance $X$, respectively.

The dominant background in $J/\psi\to\gamma\pi^+\pi^-$ comes from
$J/\psi\to\pi^+\pi^-\pi^0$. 
From Mark\,III's analysis on $J/\psi\to\pi^+\pi^-\pi^0$, described in
terms of the amplitudes representing the sequential two-body decay
process $J/\psi\to\rho\pi,~\rho\to\pi\pi$~\cite{wmd-chen},
the $\rho(770)$ is dominant, but there are also contributions from
the excited states of $\rho(770)$.
A preliminary PWA on BESII $J/\psi\to\pi^+\pi^-\pi^0$ also shows that
a complete description of the data requires not only the dominant
$\rho(770)$, but also the contributions from $1^{-}$ states $\rho(1450)$,
$\rho(1700)$ and $\rho(2150)$, as well as $3^-$ state $\rho(1690)$.
Using the branching fraction measurement of BES\,II~\cite{3pi},
which agrees with the result from the BaBar Collaboration~\cite{babar},
and the generator based on the published results of Ref.~\cite{wmd-chen},
$J/\psi\to\pi^+\pi^-\pi^0$ events are generated and
given the opposite log likelihood in the fit to cancel the background
events in the data.
The other background contributions are much smaller than
$J/\psi\to\pi^+\pi^-\pi^0$ background in the 1.0 to 2.3 GeV/$c^2$
$\pi^+\pi^-$
mass range and not considered in the partial wave analysis.

Due to the limitation of the present statistics and the complexity of
the large $\rho\pi$ background, it is difficult to cleanly distinguish
components in the high mass region. The main goal of this analysis
will be to understand the structures below 2.0 GeV/$c^2$. We use the
four states $f_2(1810)$, $f_0(2020)$, $f_2(2150)$, and $f_4(2050)$,
which are listed in PDG~\cite{PDG} in the fit with the masses and
widths fixed to those in the PDG, to describe the contribution of the
high mass states in the mass range below 2.0 GeV/$c^2$.

For the $2^{++}$ states, relative phases between different helicity
amplitudes for a single resonance are theoretically expected to be
very small~\cite{korner}.  Therefore, these relative phases are set to
zero in the final fit so as to constrain the intensities further.

After the mass and width optimization, the resulting fitted
intensities are illustrated in Figs.~3 and 4.  Angular distributions
in the whole mass range are shown in Fig.~\ref{fangcr}. Here,
$\theta_{\gamma}$ is the polar angle of the photon in the $J/\psi$
rest frame, and $\theta_{\pi}$ is the polar angle of the pion in the
$\pi^+\pi^-$ rest frame.

    \begin{figure}[htbp]
    \centerline{
    \psfig{file=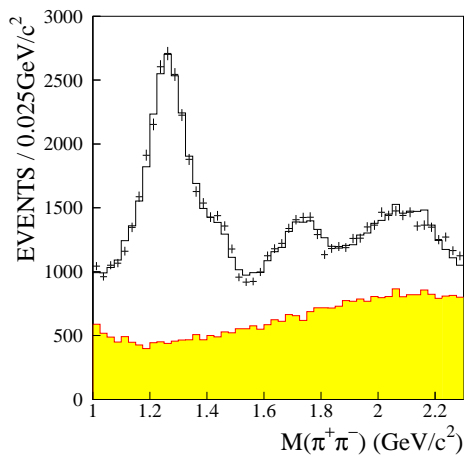,width=6.8cm,angle=0}}
    \caption{The $\pi^+\pi^-$ invariant mass distribution from
$J/\psi \to \gamma \pi^+\pi^-$. 
The crosses are data, the full histogram shows 
the maximum likelihood fit, and the shaded histogram
corresponds to the $\pi^+\pi^-\pi^0$ background.}
    \label{fm1cr}
    \end{figure}

    \begin{figure*}[htbp]
    \centerline{
    \psfig{file=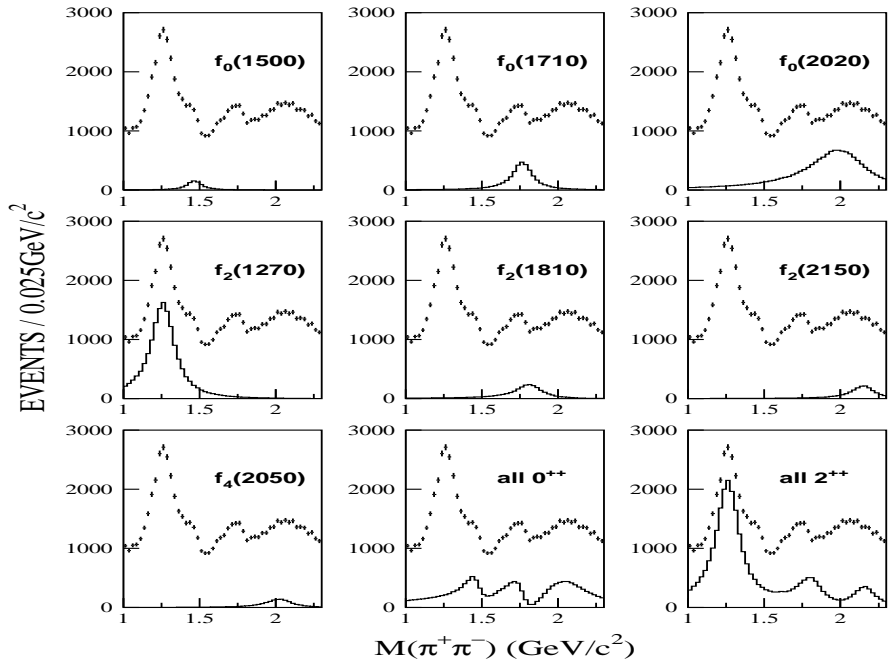,width=14.cm,height=10.cm,angle=0}}
    \caption{The mass projections of the individual components for
    $J/\psi \to \pi^+\pi^-$. The crosses are data. The complete
    $0^{++}$ and $2^{++}$ contributions are also shown, including all
    interferences. }
    \label{fm2cr}
    \end{figure*}

    \begin{figure}[hbp]
    \centerline{
    \psfig{file=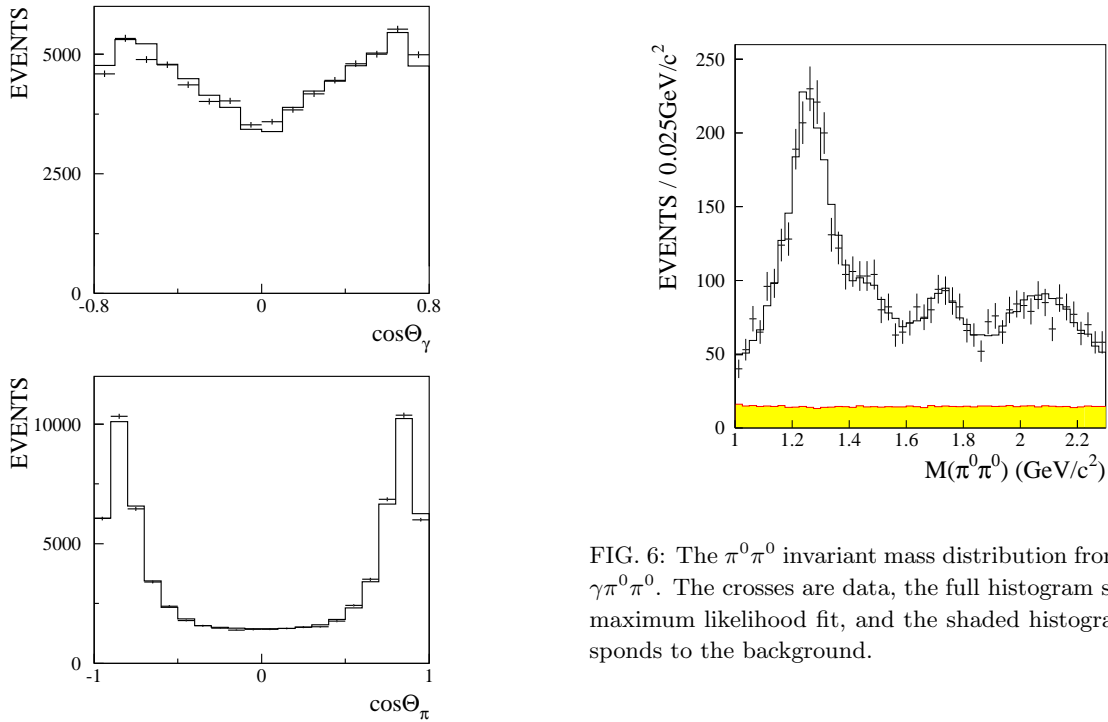,width=6.8cm,angle=0}}
    \caption{Projections in $\cos\theta_{\gamma}$ and $\cos\theta_{\pi}$ 
     for the whole mass range. The crosses are data
    ($J/\psi\to\gamma\pi^+\pi^-$ sample), and the histograms are the
    fit results.}
    \label{fangcr}
    \end{figure}

From Figs. 3 and 5, we see that the fit agrees well with data.
Fig.~\ref{fm2cr} shows the distributions of the individual components
and full $0^{++}$ and $2^{++}$ contributions including
interferences. A free fit to $f_2(1270)$ gives a fitted mass of $1262
^{+1}_{-2}$ MeV/$c^2$ and a width of $175^{+6}_{-4}$ MeV/$c^2$. The
fitted masses and widths of the $f_0(1500)$ and $f_0(1710)$ are
M$_{f_0(1500)}$ $=1466\pm6$ MeV/$c^2$,~ $\Gamma_{f_0(1500)} =
108^{+14}_{-11}$ MeV/$c^2$ and M$_{f_0(1710)}$ $=1765^{+4}_{-3}$
MeV/$c^2$,~ $\Gamma_{f_0(1710)} = 145\pm8$ MeV/$c^2$,
respectively. The branching fractions of $f_2(1270)$, $f_0(1500)$, and
$f_0(1710)$ determined by the partial wave analysis fit are ${\cal
B}(J/\psi\to\gamma f_2(1270)\to\gamma \pi^+\pi^-) =
(9.14\pm0.07)\times 10^{-4}$, ${\cal B}(J/\psi\to\ \gamma
f_0(1500)\to\gamma \pi^+\pi^-) = (6.65\pm0.21)\times 10^{-5}$, and
${\cal B}(J/\psi\to\gamma f_0(1710)\to\gamma \pi^+\pi^-) =
(2.64\pm0.04)\times 10^{-4}$, respectively. For the $f_2(1270)$, we
find the ratios of helicity amplitudes $x = 0.89 \pm 0.02$ and $y =
0.46 \pm 0.02$ with correlation factor $\rho = 0.26$, where $x=A_1/A_0$,
 $y=A_2/A_0$, $A_{0,1,2}$ correspond to the three independent production
amplitudes with helicity 0, 1 and 2. The errors here are
statistical errors. An alternative
fit is tried by replacing $f_0(1500)$ with a $2^{++}$ resonance. There
are three helicity amplitudes fitted for spin 2, while only one
amplitude for spin 0, which means the the number of degrees of freedom
is increased by 2 in the $J^P = 2^+$ case. However, the log likelihood
is worse by 108. This indicates that $f_0(1500)$ with $J^P = 0^+$ is
strongly favored.  If the $f_0(1710)$ is removed from the fit, the log
likelihood is worse by 379, which corresponds to a signal significance
much larger than $5\sigma$.

The partial wave analysis of $J/\psi\to\gamma\pi^0\pi^0$ is performed
independently. The components used are the same as those of
$J/\psi\to\gamma\pi^+\pi^-$. Due to the limited statistics, we take
the partial wave analysis fit results of $J/\psi\to\gamma\pi^0\pi^0$
as a cross-check of the ones obtained in the charged channel. The
background distribution is close to flat in the mass interval 1.0 - 2.3
GeV/$c^2$, so we use $1/p_2(m)$ multiplied by phase space to
approximately describe such a flat contribution, where the 2nd order
polynomial function $p_2(m)$ is obtained by fitting the $\pi^0\pi^0$
mass phase space distribution of $J/\psi\to\gamma\pi^0\pi^0$ Monte
Carlo simulation.

The fitted intensities as a function of $\pi^0\pi^0$ mass are
illustrated in Fig.~6. A free fit to $f_2(1270)$ gives a fitted mass
of $1261 \pm6$ MeV/$c^2$ and a width of $188^{+18}_{-16}$
MeV/$c^2$. The fitted masses and widths of the $f_0(1500)$ and
$f_0(1710)$ are M$_{f_0(1500)}$ $=1485\pm21$ MeV/$c^2$,~
$\Gamma_{f_0(1500)} = 178^{+60}_{-40}$ MeV/$c^2$ and M$_{f_0(1710)}$
$=1755\pm14$ MeV/$c^2$,~ $\Gamma_{f_0(1710)} = 155^{+30}_{-26}$
MeV/$c^2$, respectively. The errors shown here are statistical.

    \begin{figure}[htbp]
    \centerline{
    \psfig{file=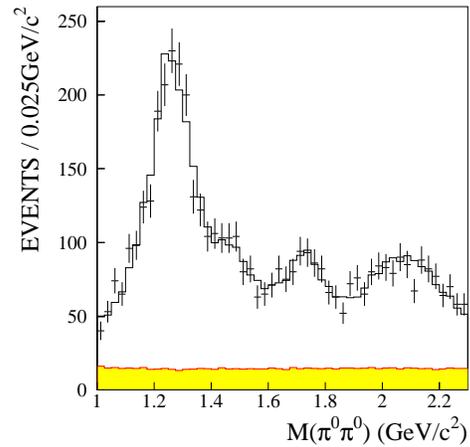,width=6.8cm,angle=0}}
    \caption{The $\pi^0\pi^0$ invariant mass distribution from 
$J/\psi \to \gamma \pi^0\pi^0$. 
The crosses are data, the full histogram shows 
the maximum likelihood fit, and the shaded histogram
corresponds to the background.}
    \label{fm1cr}
    \end{figure}

Besides the above global fit, a bin-by-bin fit is applied to
$J/\psi\to\gamma\pi^+\pi^-$ data using the method described in
Ref.~\cite{gzj}.  A strong $f_2(1270)$ is observed, and the S-wave
$\pi^+\pi^-$ mass distribution shows a large signal at $\sim$ 1.45
GeV/$c^2$, a significant signal at $\sim$ 1.75 GeV/$c^2$, and a peak
at $\sim$ 2.1 GeV/$c^2$. In general, the bin-by-bin fit gives similar
features as those of the global fit, and the results of these two fits
are approximately consistent with each other.

\vspace{-0.33cm}
\section{\boldmath Systematic errors}

The systematic errors for the partial wave analysis fit to
$J/\psi\to\gamma\pi^+\pi^-$ data are estimated by varying the masses
and widths of the $f_2(1270)$, $f_0(1500)$, and $f_0(1710)$ within the
fitted errors; varying the masses and widths of the $f_2(1810)$,
$f_0(2020)$, $f_2(2150)$, and $f_4(2050)$ within the PDG
errors~\cite{PDG}; adding a small component $f_2(1565)$; varying the
background fraction within reasonable limit and replacing the
$f_2(1810)$ with the $f_2(1950)$.  They also include the uncertainties 
in the number of $J/\psi$ events analyzed,
the efficiency of photon detection, the efficiency of MDC
tracking, and the kinematic fit. 
Different generators, one is
based on the published results of Ref.~\cite{wmd-chen} and another is
from BESII preliminary PWA results which include a dominant
$\rho(770)$ and its excited states $\rho(1450)$,
$\rho(1700)$, $\rho(2150)$ and $\rho_3(1690)$, are used for the estimation
of the background $J/\psi \to \pi^+ \pi^- \pi^0$. The PWA results
with different backgrounds agree with each other within the error.
Their difference is taken as the systematic error too.
The systematic errors in this analysis do not include all model-dependent
effects, such as the difference of using simple Breit-Wigner
formalism and K-matrix formalism.
Table~\ref{sys1} shows the summary of the systematic errors for the
global fit.

For the $f_2(1270)$, the total systematic errors are 0.10 and 0.19 for
$x$ and $y$, respectively. The correlation factor between the $x$ and $y$
systematic errors is 0.29 which is calculated with $\rho=\sum_{i}
\frac{\rho_{i} \sigma_{xi} \sigma_{yi}}{\sigma_{x}\sigma_{y}}$, where
i runs over all systematic errors.

Table~\ref{result1} shows the mass, width, and branching fraction
measurements for $f_2(1270)$, $f_0(1500)$, and $f_0(1710)$, where the
first error is statistical and the second is systematic, determined by
adding all sources in quadrature.  In order to compare the branching
fractions of $f_2(1270)$, $f_0(1500)$, and $f_0(1710)$ in
$\gamma\pi^+\pi^-$ and $\gamma\pi^0\pi^0$, we fix the mass and width
of each component in $\gamma\pi^0\pi^0$ to those of the charged
channel and re-calculate the branching fractions and estimate the
systematic errors. The results are shown in Table~\ref{result2}.  The
branching fractions determined from the two channels agree with each
other within errors after considering isospin corrections.

   \begin{table*}[bhtp]
   \begin{center}
   \caption{Estimation of systematic errors for the
           $J/\psi\to\gamma\pi^+\pi^-$ in the global fit. For the mass and
           width, what shown are the absolute errors in MeV/$c^2$. For 
           the branching fraction ${\cal B}$ and the ratios of helicity 
           amplitudes, $x$ and $y$, the listed are the relative errors. 
           $\rho$ is the correlation factor between $x$ and $y$. }
   \begin{tabular}{l|c|c|c} \hline\hline
     &$f_2(1270)$ &$f_0(1500)$ &$f_0(1710)$\\ \hline
  &\,~~~M~~~~$\Gamma$~~~~~${\cal B}$ (\%)~~~~$x$ (\%)~~~~$y$(\%)~~~~~$\rho$~~&\,~~~M~~~~$\Gamma$\,~~~${\cal B}$ (\%)&\,~~~M~~~~$\Gamma$\,~~~~${\cal B}$ (\%) \\ \hline
   M and $\Gamma$ of $f_2(1270)$
&~~~~~~~~~~~~~~~~~~1.4~~~~~~~~0.2~~~~~~~2.0~~~~~0.26&~0~~~~~0~~~~~2.3&~0~~~~~0~~~~~~0.1  \\ \hline
   M and $\Gamma$ of $f_0(1500)$ 
&~~~~1~~~~~0~~~~~~0.1~~~~~~~~0.1~~~~~~~2.2~~~~~0.26 &~~~~~~~~~~~~~10.1&~0~~~~~1~~~~~~1.9  \\ \hline
   M and $\Gamma$ of $f_0(1710)$ 
&~~~~0~~~~~0~~~~~~0.1~~~~~~~~0.5~~~~~~~0.6~~~~~0.26 &~0~~~~~1~~~~~2.6&~~~~~~~~~~~~~~~8.9 \\ \hline
   M and $\Gamma$ of $f_2(1810)$ 
&~~~~0~~~~~1~~~~~~1.4~~~~~~~~2.9~~~~~~~0.5~~~~~0.26 &~1~~~~~7~~~~~4.2&~3~~~~15~~~~~~6.9 \\ \hline
   M and $\Gamma$ of $f_0(2020)$ 
&~~~~1~~~~~3~~~~~~1.0~~~~~~~~1.6~~~~~~~1.2~~~~~0.27 &~6~~~~~2~~~~~3.0&~5~~~~10~~~~~~5.8 \\ \hline
   M and $\Gamma$ of $f_2(2150)$ 
&~~~~1~~~~~2~~~~~~1.2~~~~~~~~1.2~~~~~~~0.4~~~~~0.26 &~3~~~~~0~~~~~3.2&~1~~~~~3~~~~~~3.3  \\ \hline
   M and $\Gamma$ of $f_4(2050)$ 
&~~~~0~~~~~1~~~~~~0.1~~~~~~~~0.4~~~~~~~0.7~~~~~0.26 &~1~~~~~2~~~~~0.4&~0~~~~~0~~~~~~0.7  \\ \hline
   add $f_2(1565)$          
&~~~~2~~~~~5~~~~~~0.2~~~~~~~~6.9~~~~~~34.6~~~~~0.35 &~3~~~~~7~~~~18.6&~3~~~~10~~~~~~2.9  \\ \hline
 MDC tracking and kinematic fit     
&~~~~4~~~~~4~~~~~~3.5~~~~~~~~6.2~~~~~~12.8~~~~~0.38 &14~~~~18~~~~31.0&~8~~~~~1~~~~~13.4  \\ \hline
    background fraction                    
&~~~~3~~~~~2~~~~~~1.8~~~~~~~~3.4~~~~~~~4.3~~~~~0.33 &~4~~~~~2~~~~13.9&~3~~~~10~~~~~~7.8 \\ \hline
  background with different generator
&~~~~3~~~~~3~~~~~11.8~~~~~~~~1.7~~~~~~19.5~~~~~0.41 &~2~~~~13~~~~14.6&~5~~~~~7~~~~~~~9.0 \\ \hline 
  replace $f_2(1810)$ with $f_2(1950)$ 
&~~~~4~~~~~5~~~~~~8.7~~~~~~~~~2.1~~~~~~~4.4~~~~~0.27 &12~~~~~4~~~~11.0&~5~~~~65~~~~~16.4 \\ \hline
  $\delta_{N_{J/\psi}}$          
& 4.7~~~~~~~~~~~~~~~ &~~~~~~~~~~~~~~~4.7&~~~~~~~~~~~~~~~~4.7\\ \hline
  Detection efficiency 
 of photon        
& 2.0~~~~~~~~~~~~~~~ &~~~~~~~~~~~~~~~2.0&~~~~~~~~~~~~~~~~2.0\\ \hline      
Total     Systematic error
&~~~~8~~~~10~~~~~16.2~~~~~~~10.9~~~~~~42.3~~~~~0.29 &20~~~~25~~~~44.8&13~~~~69~~~~~28.3 \\ \hline\hline
   \end{tabular}
   \label{sys1}
   \end{center}
%
%
   \caption{ Fit results for $J/\psi\to\gamma\pi^+\pi^-$.
     The first error is statistical, and the second is systematic.}
   \begin{center}
   \begin{tabular}{ l  c c r  } \hline\hline
      & \multicolumn{3}{c}{$J/\psi\to\gamma X,~X\to\pi^+\pi^-$}  \\ \hline
       &Mass~(MeV/$c^2$) & $\Gamma$~(MeV/$c^2$) & ${\cal B}$~($\times 10^{-4}$) \\ \hline
$f_2(1270)$~~~~&$1262^{+1}_{-2}\pm8 $ &$ 175^{+6}_{-4}\pm10 $
&~~$9.14\pm0.07\pm1.48 $ \\ \hline
$f_0(1500)$ &~~~$1466\pm6\pm20 $~~~&$~~~108^{+14}_{-11}\pm25
$~~~&~~~$0.67\pm0.02\pm0.30 $\\ \hline
$f_0(1710)$ &$1765^{+4}_{-3}\pm13 $ &$ 145\pm8\pm69 $  
&$2.64\pm0.04\pm0.75 $ \\ \hline\hline
   \end{tabular}
   \label{result1}
   \end{center}
%
%
   \caption{ The branching fraction measurements of
    $J/\psi\to\gamma\pi^0\pi^0$, where the masses and widths of the
    resonances are fixed to the values determined from
    $J/\psi\to\gamma\pi^+\pi^-$. The first error is statistical, and
    the second is systematic.}
   \begin{center}
   \begin{tabular}{l  c r  } \hline\hline
       & \multicolumn{2}{c}{$J/\psi\to\gamma X,~X\to\pi^0\pi^0$}\\ \hline
 &Mass~(MeV/$c^2$)~~~~$\Gamma$~(MeV/$c^2$) & ${\cal B}$~($\times 10^{-4}$) \\ \hline
$f_2(1270)$~~~&~~~same as charged channel~~~&~~~$4.00\pm0.09\pm0.58 $\\ \hline
$f_0(1500)$~~~&~same as charged channel &$0.34\pm0.03\pm0.15 $ \\ \hline
$f_0(1710)$~~~&~same as charged channel &$1.33\pm0.05\pm0.88 $\\ \hline\hline
   \end{tabular}
   \label{result2}
   \end{center}
   \end{table*}

\section{\boldmath discussion}

The measured mass of the $f_2(1270)$, $1262^{+1}_{-2}\pm8$ MeV/$c^2$,
is lower than the PDG value, and the branching fraction of
$J/\psi\to\gamma f_2(1270),~f_2(1270)\to\pi^+\pi^-$ is a bit higher
than the PDG value~\cite{PDG}. A fit with the PDG mass and width is
visibly poorer, and the log likelihood is worse than the optimum fit
by 44.  In this analysis, the S-wave contribution on the high mass
shoulder of the large $f_2(1270)$ peak is well separated, which may
explain this mass difference.  The ratios of the helicity amplitudes
of the $f_2(1270)$ from the present analysis are $x=0.89 \pm 0.02 \pm
0.10$ and $y=0.46 \pm 0.02 \pm 0.19$ with correlation
$\rho_{stat}=0.26$ and $\rho_{sys}=0.29$. The values of $x$ and $y$ are in
agreement with predictions~\cite{korner,krammer} within the errors.
Comparing to the results determined by DM2~\cite{dm2},
Mark\,III~\cite{mark3}, and Crystal Ball~\cite{cball2}, there is a
difference in the value of y. The main reason for this difference may be
that we consider influences from other states in the 1.0 to 2.3
GeV/$c^2$ $\pi \pi$ mass range, while previous analyses by DM2 and
Mark\,III ignored these and only considered the $f_2(1270)$
in the 1.15-1.4 GeV/$c^2$ $\pi \pi$ mass range.
 
The most remarkable feature of the above results is that three
$0^{++}$ states at the $\pi\pi$ mass $\sim$ 1.45, 1.75, and 2.1
GeV/$c^2$ are observed from the partial wave analysis. For the high
mass state at $\sim$ 2.1 GeV/$c^2$, we use the $f_0(2020)$, which is
listed in the PDG, to describe it. No further efforts are made on the
measurements of its resonant parameters due to the difficulties
described in Section IV.

The lower $0^{++}$ state peaks at a mass of $1466\pm 6\pm 20$
MeV/$c^2$ with a width of $108{^{+14}_{-11}}\pm 25$ MeV/$c^2$, which
is consistent with the scalar glueball candidate, $f_0(1500)$.  Spin 0
is strongly preferred over spin 2.  Therefore we interpret the small
but definite shoulder on the high mass side of the $f_2(1270)$ in the
$\pi^+\pi^-$ mass distribution as originating from the
$f_0(1500)$, which interferes with nearby resonances in the partial
wave analysis. However, due to the uncertainties of the mass and width
determinations and large interferences between the S-wave states, the
existence of the $f_0(1370)$ in $J/\psi\to\gamma\pi\pi$ cannot be
excluded by present data.

Strong production of the $f_0(1710)$ was observed in the partial wave
analysis of $J/\psi\to\gamma K \bar{K}$, with a mass of
$1740\pm4^{+10}_{-25}$ MeV/$c^2$, a width of $166^{+5+15}_{-8-10}$
MeV/$c^2$, and a branching fraction of $J/\psi\to\gamma
f_0(1710),~f_0(1710)\to K \bar{K}$ of
$(9.62\pm0.29^{+2.11+2.81}_{-1.86-0.00})\times 10^{-4}$~\cite{gzj}.
Interpreting the $0^{++}$ state in the mass region $\sim$ 1.75 GeV/$c^2$
as coming from the $f_0(1710)$ and using the
branching fraction of $f_0(1710)\to\pi\pi$ determined from
$f_0(1710)\to\pi^+\pi^-$ after isospin correction and the branching
fraction of $f_0(1710)\to K\bar K$ in Ref.~\cite{gzj}, we obtain the
ratio of $\pi\pi$ to $K\bar K$ branching fractions for the $f_0(1710)$ as
$$
\frac{\Gamma(f_0(1710)\to\pi\pi)}{\Gamma(f_0(1710)\to K\bar{K})}
=0.41^{+0.11}_{-0.17}.
$$
The ratio is consistent with the PDG value ($0.200\pm0.024\pm0.036$)
~\cite{PDG} within the errors. An alternative interpretation for this
$0^{++}$ state is the $f_0(1790)$.  Data on $J/\psi \to \phi \pi ^+\pi
^-$ and $\phi K^+K^-$ show a definite peak in $\pi ^+\pi ^-$ at 1790
MeV/$c^2$ but no significant signal in $K^+K^-$~\cite{phipp}. If the
mass and width of the $0^{++}$ state are fixed to 1790 MeV/$c^2$ and
270 MeV/$c^2$ found in~\cite{phipp}, the log likelihood is worse by
47. This $0^{++}$ state may also be a superposition of $f_0(1710)$ and
$f_0(1790)$.
 
Due to the uncertainty of the high mass region, we do an alternative fit
removing the $f_4(2050)$ and re-optimizing the masses and widths of
$f_2(1270)$, $f_0(1500)$, and $f_0(1710)$ for the
$J/\psi\to\gamma\pi^+\pi^-$ sample. The log likelihood is worse by
160.  The fit gives a $f_2(1270)$ mass of $1259\pm2\pm6$ MeV/$c^2$ and
a width of $175 ^{+4}_{-5}\pm8$ MeV/$c^2$.  The measured masses and
widths of the $f_0(1500)$ and $f_0(1710)$ are M$_{f_0(1500)}$
$=1466\pm6\pm17$ MeV/$c^2$, $\Gamma_{f_0(1500)} =
118^{+14}_{-15}\pm28$ MeV/$c^2$ and M$_{f_0(1710)}$
$=1768^{+5}_{-4}\pm15$ MeV/$c^2$,~ $\Gamma_{f_0(1710)} =
112^{+10}_{-8}\pm52$ MeV/$c^2$, respectively. The branching fractions
of the $f_2(1270)$, $f_0(1500)$, and $f_0(1710)$ are ${\cal
B}(J/\psi\to f_2(1270)\to\gamma \pi^+\pi^-) = (9.04\pm0.07\pm0.89)
\times 10^{-4}$, ${\cal B}(J/\psi\to \gamma f_0(1500)\to\gamma
\pi^+\pi^-) = (0.68\pm0.02\pm0.28)\times 10^{-4}$, and ${\cal
B}(J/\psi\to\gamma f_0(1710)\to\gamma \pi^+\pi^-) =
(1.40\pm0.03\pm0.55)\times 10^{-4}$, respectively.
 
The light-meson spectroscopy of scalar states in the mass range 1-2
GeV/$c^2$, which has long been a source of controversy, is still very
complicated.  Overlapping states interfere with each other differently
in different production and decay channels. More experimental data are
needed to clarify the properties of these scalar states.

\section{\boldmath summary}

  In summary, the partial wave analyses of $J/\psi\to\gamma\pi^+\pi^-$
  and $J/\psi\to\gamma\pi^0\pi^0$ using 58M $J/\psi$ events of 
  BES\,II show strong production of $f_2(1270)$ and 
  evidence for two $0^{++}$ states in the 1.45 and 1.75 GeV/$c^2$ mass
  regions. For the $f_2(1270)$, the branching ratio determined 
  by the partial wave analysis fit is 
  ${\cal B}(J/\psi\to \gamma f_2(1270)\to\gamma \pi^+\pi^-) =
  (9.14\pm0.07\pm1.48)\times 10^{-4}$. The ratios of the helicity amplitudes
  of the $f_2(1270)$ are determined to be $x=0.89\pm0.02\pm0.10$ and
  $y=0.46\pm0.02\pm0.19$ with correlations $\rho_{stat}=0.26$ and
  $\rho_{sys}=0.29$.
  The $f_0(1500)$ has a mass of $1466\pm
  6\pm 20$ MeV/$c^2$, a width of $108{^{+14}_{-11}}\pm 25$ MeV/$c^2$, and 
  a branching fraction 
  ${\cal B}(J/\psi\to \gamma f_0(1500)\to\gamma \pi^+\pi^-) =
  (0.67\pm0.02\pm0.30)\times 10^{-4}$.
  The $0^{++}$ state in the $\sim$1.75 GeV/$c^2$ mass region has a mass of
  $1765{^{+4}_{-3}}\pm 13$ MeV/$c^2$ and a width of $145\pm 8 \pm 69$ MeV/$c^2$.
  If this $0^{++}$ state is interpreted as coming from $f_0(1710)$,  
  the ratio of the $\pi\pi$ to $K\bar K$ branching fractions is
  $0.41^{+0.11}_{-0.17}$. This may help in understanding the
  properties of $f_0(1500)$ and $f_0(1710)$.

\section{\boldmath Acknowledgments}

The BES collaboration thanks the staff of BEPC and computing center
for their hard efforts.  We wish to thank Prof. David Bugg for
contributions to the early stage of this analysis.  This
work is supported in part by the National Natural Science Foundation
of China under contracts Nos. 10491300, 10225524, 10225525, 10425523,
the Chinese Academy of Sciences under contract No. KJ 95T-03, the 100
Talents Program of CAS under Contract Nos. U-11, U-24, U-25, and the
Knowledge Innovation Project of CAS under Contract Nos. KJCX2-SW-N10, U-602, U-34
(IHEP), the National Natural Science Foundation of China under
Contract No. 10225522 (Tsinghua University), and the Department of
Energy under Contract No.DE-FG02-04ER41291 (U Hawaii).

\end{document}